\documentclass[final,1p,times]{elsarticle}

\usepackage{amsmath,amssymb,amsfonts}
\usepackage{graphicx}
\usepackage{adjustbox}
\usepackage{array}
\newcolumntype{L}{>{\centering\arraybackslash}p{2.5cm}}

\journal{Physics Letters B}

\begin{document}

\begin{frontmatter}



\title{Neutron oscillations for solving neutron lifetime and dark matter puzzles}


\author{Wanpeng Tan}
\ead{wtan@nd.edu}
\address{Department of Physics, Institute for Structure and Nuclear Astrophysics (ISNAP), and Joint Institute for Nuclear Astrophysics - Center for the Evolution of Elements (JINA-CEE), University of Notre Dame, Notre Dame, Indiana 46556, USA}
\address{\textit{This article is registered under preprint number: /hep-ph/1902.01837}}

\begin{abstract}
A model of $n-n'$ (neutron-mirror neutron) oscillations is proposed under the framework of the mirror matter theory with slightly broken mirror symmetry. It resolves the neutron lifetime discrepancy, i.e., the 1\% difference in neutron lifetime between measurements from ``beam'' and ``bottle'' experiments. In consideration of the early universe evolution, the $n-n'$ mass difference is determined to be about $2\times 10^{-6}$ eV/c$^2$ with the $n-n'$ mixing strength of about $2\times 10^{-5}$. The picture of how the mirror-to-ordinary matter density ratio is evolved in the early universe into the observed dark-to-baryon matter density ratio of about 5.4 is presented. Reanalysis of previous data and new experiments that can be carried out under current technology are discussed and recommended to test this proposed model. Other consequences of the model on astrophysics and possible oscillations of other neutral particles are discussed as well.

\end{abstract}


\end{frontmatter}



Free neutrons with a lifetime of about 15 minutes are known to undergo $\beta$ decay via $n\rightarrow p + e^- + \bar{\nu}_e$ due to the weak force. There has been much experimental effort over the past decades for measuring the lifetime using two different techniques. The ``beam'' approach is to measure the neutron flux from a cold neutron beam after it going through a region where the emitted protons are detected \cite{nico2005,yue2013}. It measures directly the $\beta$ decay rate as far as other hidden neutron-disappearing processes are on the level of $10^{-3}$ or below. This approach typically gives a neutron lifetime of about 888 seconds. On the other hand, the ``bottle'' experiments store ultra-cold neutrons (UCN) confined by the gravitational force in a material or magnetic trap \cite{pattie2018,serebrov2018,arzumanov2015}. By measuring the neutron loss rate in the trap this method typically presents a neutron lifetime of about 880 seconds. Note that any other unknown loss processes in the trap will contribute to the measured lifetime and make it appear shorter. Another different approach using a magnetic storage ring \cite{paul1989} provides similar results as the ``bottle'' method. The 1\% difference between the results of the two approaches becomes more severe recently with the most precise measurements of $887.7\pm 1.2 (stat)\pm 1.9 (sys)$ s (``beam'') \cite{yue2013} and $877.7 \pm 0.7 (stat) +0.4/-0.2 (sys)$ s (``bottle'') \cite{pattie2018}.

Meanwhile, various theoretical studies on resolving the 1\% neutron lifetime discrepancy have been carried out. Searching physics beyond the standard model makes the idea of $n-\bar{n}$ oscillations intriguing. However, an early experiment set a very strict constraint on the oscillation time scale $\tau_{n\bar{n}} > 0.86 \times10^8$ s \cite{baldo-ceolin1994} making it unlikely to settle the issue. A recent attempt to consider neutrons that decay to particles in the dark sector showed an interesting decay channel of $n \rightarrow \chi + \gamma$ with constraints of $937.900$ MeV $< m_{\chi} < 938.783$ MeV for the dark particle mass and $0.782$ MeV $< E_{\gamma} < 1.664$ MeV for the photon energy \cite{fornal2018}. Unfortunately, such a possibility was dismissed shortly by an experiment \cite{tang2018} and a similar channel of $n \rightarrow \chi + e^+ + e^-$ was excluded as well \cite{ucnacollaboration2018}. By introducing a six-quark coupling in the mirror matter theory for the $n$ and $n'$ interaction of $\delta m \sim 10^{-15}$ eV with a large mass cutoff at $M \sim 10$ TeV, Berezhiani and Bento proposed a possible $n-n'$ oscillation mechanism with a time scale of $\tau \sim 1$ s \cite{berezhiani2006}. Later on, such oscillations were refuted experimentally with a much higher constraint of $\tau \geq 448$ s \cite{serebrov2009,ban2007,altarev2009,serebrov2008}. Despite all these efforts over the years the neutron lifetime puzzle still eludes explanation. More recent papers that have come to my awareness after the preparation of this work suggest other interesting ideas on either neutron dark decays or $n-n'$ oscillations \cite{karananas2018,bringmann2019,goldman2019,berezhiani2019a}. In particular, a $n-n'$ oscillation model was proposed using a six-quark coupling and a small $n-n'$ mass splitting of $10^{-7}$ eV \cite{berezhiani2019a} where, like many other studies, the ``bottle'' lifetime is favored.

In this paper, a new mechanism of $n-n'$ oscillations will be proposed. The new model can explain the observed difference of neutron lifetime measurements without harming other known physics with a $n-n'$ mixing strength of $2\times 10^{-5}$. Considering the thermal history of the early universe and big bang nucleosynthesis (BBN), we will show the mass difference of the $n-n'$ doublet to be about $2\times 10^{-6}$ eV/c$^2$ under the framework of the mirror matter theory with slightly broken mirror symmetry and no explicit cross-sector interaction. How the early universe evolved with both sectors forming the observed dark matter to baryon matter ratio $\Omega_{dark}/\Omega_B = 5.4$ will be demonstrated under this model. Possible experimental tests to confirm or refute this model will be discussed along the way and in the end.

The idea that there may exist mirror particles that compensate the parity violation of ordinary particles in the universe was first conceived by Lee and Young in their seminal paper on parity violation \cite{lee1956}. The idea has been developed into theories of a parallel world of mirror particles that is an exact mirrored copy of our ordinary world and the two worlds can only interact with each other gravitationally \cite{kobzarev1966,blinnikov1982,blinnikov1983,kolb1985,khlopov1991,foot2004,berezhiani2004,okun2007}. 
Such a mirror matter theory has appealing theoretical features. For example, it can be embedded in the $E_8\otimes E_{8'}$ superstring theory \cite{green1984,gross1985,kolb1985} and it can also be a natural extension of recently developed twin Higgs models \cite{chacko2006,barbieri2005} that protect the Higgs mass from quadratic divergences and hence solve the hierarchy or fine-tuning problem. The mirror symmetry or twin Higgs mechanism is particularly intriguing as the Large Hadron Collider has found no evidence of supersymmetry so far and we may not need supersymmetry, at least not below energies of 10 TeV.

For simplicity, one can consider a gauge symmetry $G\otimes G'$ for both sectors of ordinary and mirror particles, where the standard model symmetry $G = SU(3)_c \otimes SU(2)_L \otimes U(1)_Y$ and the mirror counterpart $G'=SU(3)'_c \otimes SU(2)'_R \otimes U(1)'_Y$. The two parallel worlds share nothing but the same gravity. Very importantly, we assume that the mirror symmetry $\mathcal{M}(G \leftrightarrow G')$ is spontaneously broken by the Higgs vacuum, i.e., $<\phi> \neq <\phi'>$, although very slightly (e.g., on a relative breaking scale of $10^{-15} \text{--} 10^{-14}$ in this work). Mass of a fermion particle $\psi$ will be obtained via the Yukawa term of the Lagrangian coupled to the Higgs field $\phi$ owing to the broken symmetry,
\begin{equation}
\mathcal{L}_{Yukawa} = -Y^{\alpha\beta}\bar{\psi}_{L\alpha} \psi_{R\beta} \phi_{\beta} + h.c.
\end{equation}
where $\alpha$ and $\beta$ are the mirror indices of 1 and 2 of the two sectors. Note that this mirror mixing is similar to the family mixing for quarks and neutrinos in the standard model and the basis of mass eigenstates is not the same as that of mirror eigenstates. Therefore, like the CKM and PMNS matrices, a unitary mirror mixing operator is defined as follows,
\begin{equation}
U = 
\begin{pmatrix}
\cos \theta_m & \sin \theta_m \\
-\sin \theta_m & \cos \theta_m
\end{pmatrix}
\end{equation}
which transforms between the two bases with a mixing angle of $\theta_m$.

This broken mirror symmetry then naturally leads to the oscillations of neutral particles due to a mass difference. Similar to the ordinary neutrino oscillation, we can find the probability of non-relativistic $n-n'$ oscillations in free space,
\begin{equation}\label{eq_prob}
P_{nn'}(t) = \sin^2(2\theta) \sin^2(\frac{1}{2}\Delta_{nn'} t)
\end{equation}
where $\theta$ is the $n-n'$ mixing angle and $\sin^2(2\theta)$ denotes the mixing strength, $t$ is the propagation time, and $\Delta_{nn'} = m_{n} - m_{n'}$ is the small mass difference. Note that such oscillations do not affect the stability of nuclei with bound neutrons owing to energy conservation. From now on, natural units ($\hbar=c=1$) are used for simplicity and quantities of the mirror particles will be marked by $'$ to distinguish from those of the ordinary particles. For $t \ll \tau_{\beta} \approx 888$ s, the neutron $\beta$ decay factor of $\exp(-t/\tau_{\beta})$ is omitted in Eq. (\ref{eq_prob}).

If neutrons travel in a magnetic field $B$, Eq. (\ref{eq_prob}) will generally be modified by a medium effect due to the effective potential contributed to the Hamiltonian from the field \cite{berezhiani2019a,tan2019a,tan2019d}. However, such an effect is negligible \cite{tan2019a,tan2019d} if $\mu B \ll \Delta_{nn'}$ where $\mu=|\mu_n| \approx 6\times 10^{-8}$ eV/T is the absolute neutron magnetic moment. As shown below, the magnetic fields in the lifetime measurements are low enough in comparison with $\Delta_{nn'}$ and therefore Eq. (\ref{eq_prob}) still holds for the discussions below. Similar effect of $\mu' B'$ has to be considered as well if a mirror magnetic field exists at the same time.

The energy of a trapped UCN is typically less than $10^{-7}$ eV and its mean free flight time $\tau_f$ is on the order of $0.1$ s in a ``bottle'' experiment setup. Each scattering of UCN (e.g., from the trap walls) will collapse its wave function into a mirror eigenstate with a $n-n'$ transition probability $P_{nn'}(\tau_f)$ determined as in Eq. (\ref{eq_prob}). For a unit holding time in the trap, the number of such collisions will be $1/\tau_f$. Therefore, the transition rate of $n-n'$ for the trapped UCN is simply,
\begin{equation}\label{eq_prob2}
\lambda_{nn'} = \frac{1}{\tau_f}\sin^2(2\theta) \sin^2(\frac{1}{2}\Delta_{nn'} \tau_f).
\end{equation}
A more careful treatment of the $n$ and $n'$ wave function was carried out in Ref. \cite{kerbikov2008} as the trap walls do not exist for $n'$. Nonetheless, the result is the same as Eq. (\ref{eq_prob2}).

For the ``bottle'' experiments, the magnetic field of the UCN trap varied from as low as $B\approx2$ nT up to 10 mT (including ambient Earth's magnetic field of about $50\mu$T) \cite{ban2007,altarev2009,serebrov2008,serebrov2009,pattie2018} corresponding to an energy shift of $1.2\times 10^{-16} - 6\times 10^{-10}$ eV. If the $n-n'$ mass difference is large enough ($\gg 6\times 10^{-10}$ eV), the medium effect from the magnetic field will be negligible and meanwhile $\frac{1}{2}\Delta_{nn'} \tau_f \gg 1$, i.e., the propagation factor of Eq. (\ref{eq_prob2}) will simply be the mean value of 1/2. However, if the $n-n'$ mass difference is even greater than the energy (about $10^{-7}$ eV) of the trapped UCN, the propagation factor of Eq. (\ref{eq_prob2}) has to have its sine phase modified \cite{kerbikov2008} but its average is still 1/2. So under the assumption of $\Delta_{nn'} \gg 6\times 10^{-10}$ eV, we can obtain the transition rate of $n-n'$ for ``bottle'' experiments,
\begin{equation}\label{eq_prob3}
\lambda_{nn'}(\text{bottle}) = \frac{1}{2\tau_f}\sin^2(2\theta)
\end{equation}
which depends only on the mean free flight time $\tau_f$ and the mixing strength constant $\sin^2(2\theta)$ for $n-n'$ to be determined later.

There was actually strong evidence to support Eq. (\ref{eq_prob3}) from an early ``bottle'' experiment \cite{mampe1989}. They developed a novel technique with an adjustable Fomblin-coated UCN storage vessel \cite{bates1983,ageron1986} to determine the lifetime by extrapolating to the ideal condition of zero wall collisions. By varying the size of the vessel, they conducted a number of runs with effectively varied mean free flight time for UCN. Then they fit the data to an equation that is essentially the same as Eq. (\ref{eq_prob3}) and obtained the lifetime of $887.6 \pm 1.1$ s which is almost identical to the best ``beam'' measurement \cite{yue2013}. The remarkable fit in Fig. 2 of the paper \cite{mampe1989} essentially claims a hidden constant just like the $n-n'$ mixing strength $\sin^2(2\theta)$. Unfortunately, the dominating idea for the mythical loss from wall collisions was to blame the imperfect wall surface. And they were not confident of large corrections they had to apply so they changed the measured error bar from $\pm1.1$ to $\pm 3$ s. Nevertheless, the $n-n'$ mixing strength of about $2\times10^{-5}$ can be inferred from their work and the mean UCN loss per bounce on the Fomblin surface they measured essentially set an upper limit on the $n-n'$ mixing strength of $\sin^2(2\theta) \leq 4\times10^{-5}$.

As for the most recent ``bottle'' result by the UCN$\tau$ collaboration \cite{pattie2018} with a magnetic trap, neutrons are confined by magnetic fields and gravity and therefore it does not suffer the type of UCN losses from walls as in material trap experiments. However, its measured neutron lifetime is still about 1\% lower than the ``beam'' results. Taking into account the geometry of their trap, it is reasonable to estimate $\tau_f \sim 0.8$ s in their experiment. Together with the $n-n'$ mixing strength of $2\times10^{-5}$ as discussed above, the lifetime discrepancy is perfectly resolved using Eq. (\ref{eq_prob3}). Assuming that in the extreme case UCN is prepared at a much higher temperature than its kinetic energy in the trap, we can estimate a lower limit for the mixing strength $\sin^2(2\theta) \geq 8\times10^{-6}$. Under the new $n-n'$ oscillation model, magnetic traps with different sizes or effectively different mean free flight times will give different apparent lifetime values that can only be reconciled by Eq. (\ref{eq_prob3}). Future experiments with more of this type of traps will present a very strict test of this $n-n'$ oscillation model.

As a matter of fact, other ``bottle'' measurements with less precise magnetic traps have already indicated such discrepancies due to different trap sizes or mean free flight times \cite{dzhosyuk2005,leung2016,ezhov2018}. For example, a neutron storage lifetime of $874.6 \pm 1.7$ s was reported with a magnetic trap operated at the UCN facility of the Institut Laue–Langevin (ILL), France \cite{ezhov2018}. Assuming the discrepancy is all from $n-n'$ oscillations, we can obtain the mean neutron velocity of $29$ cm/s by using a mean free path of 17 cm according to the geometry of their setup. At the same facility (ILL), a very different magnetic trap (HOPE) was used to measure the neutron lifetime as well \cite{leung2016}. The HOPE trap was designed with a very thin cylindrical volume and a movable UCN remover rod at the top was used to measure the lifetime at two different heights of 65 and 80 cm, respectively. Due to the design, the neutron mean free path is essentially the same (i.e., the diameter of 9 cm) for both heights. The similar neutron energy distribution can be safely assumed for both Ref. \cite{ezhov2018} and Ref. \cite{leung2016} at the same facility. Therefore the mean neutron velocity in the HOPE experiment should be a little more than double that in Ref. \cite{ezhov2018} (i.e., 60 cm/s for height of 65 cm and 70 cm/s for 80 cm) as the maximum neutron energy in the HOPE setup is more than four to five times more. Considering $n-n'$ oscillations, the resulting lifetime within a few seconds agrees very well with the measured values of 835 s (at 65 cm) and 824 s (at 80 cm) although very large errors were applied in Ref. \cite{leung2016}. An earlier measurement at NIST \cite{dzhosyuk2005} used a magnetic trap that was very similar in geometry to the HOPE trap and a very similar lifetime of 833 s was obtained although quoted with large errors.

Now one can take a look at ``beam'' experiments in which neutrons don't bounce around until they hit the flux-monitoring detector in the end. Therefore one can consider it like traveling in free space as described in Eq. (\ref{eq_prob}). The flight time of $t \sim 10^{-3}$ s can be calculated for a flight path of 1 m and energy of 0.0034 eV \cite{nico2005}. ``Beam'' experiments typically apply high magnetic field of several Teslas (e.g. $B=4.6$ T \cite{nico2005}) to confine and extract emitted protons. Assuming that $\Delta_{nn'} \gg 3\times 10^{-7}$ eV, we can neglect the magnetic medium effect again for $B<5$ T and it makes the last factor of Eq. (\ref{eq_prob}) averaged to 1/2 as well. Therefore, the $n-n'$ transition probability is as follows,
\begin{equation}\label{eq_prob4}
P_{nn'}(\text{beam}) = \frac{1}{2}\sin^2(2\theta)
\end{equation}
which is on the order of $10^{-5}$, i.e., smaller than the best experimental precision by two orders of magnitude and basically not detectable in a ``beam'' experiment. Therefore, $n-n'$ oscillations do not affect the beta decay rate or $\tau_\beta$ measured in ``beam'' experiments. If $\Delta_{nn'} \sim 3\times 10^{-7}$ eV, i.e., $\mu B \sim \Delta_{nn'}$, the ``beam'' experiments could present a resonant $n-n'$ oscillation behavior \cite{berezhiani2019a}. As far as the mass splitting parameter is more than 10\% away from the resonant value, a ``beam'' experiment will not observe the effect on $\tau_\beta$. If $\mu B \gg \Delta_{nn'}$, the medium effect will greatly suppress the oscillation probability of Eq. (\ref{eq_prob4}) \cite{tan2019a,tan2019d} making the effect even smaller for ``beam'' experiments.

Here it is worth pointing out that we don't need the mirror-symmetry framework just to resolve the neutron lifetime discrepancy. The only assumptions for it to work are the mixing mechanism via some spontaneously broken symmetry and the mass difference should be $\gg 6\times 10^{-10}$ eV avoiding the resonant region of $\sim 3\times 10^{-7}$ eV for ``beam'' experiments . However, the mirror symmetry theory naturally presents a very elegant solution if not the best. In addition, to further constrain the $n-n'$ mass difference or better yet to nail it down, the mirror-symmetry theory need to be applied to the thermal evolution of the early universe which will be discussed below. As a motivation bonus, much richer physics can be studied under this model, for example, possible oscillations of other neutral particles and its impact on astrophysical environments.

In the first second of the Big Bang after protons and neutrons are formed from quarks, the age of the universe can be parameterized for temperatures between $10^{12}$ K ($\sim 100$ MeV) and $10^{10}$ K ($\sim 1$ MeV) as \cite{weinberg1972},
\begin{equation}\label{eq_age}
t=3.07/(\sqrt{g_*(T)} T_{10}^2) \text{[sec]} \sim \frac{1}{ T_{10}^2} \text{[sec]}
\end{equation}
where $T_{10}$ is the temperature in unit of $10^{10}$ K and $g_*(T)$ is the effective number of relativistic degrees of freedom at the given temperature $T$, which is about $10-17$ for this temperature range for one sector (the contribution from the other sector is negligible if its temperature is much lower as discussed below). As pions and muons are quickly annihilated in this temperature range, their contributions here and possible pion-neutron interactions that affect discussions below are omitted for simplicity. See Ref. \cite{bringmann2019} on the effect of pion-neutron interactions under a different neutron oscillation mechanism.

Once formed at temperature just above $10^{12}$ K, protons and neutrons are in thermal equilibrium with a 1:1 ratio by interacting with electrons, positrons, and neutrinos. They each consist of half of the baryon content because the Q-value or the mass difference between proton and neutron (1.293 MeV) is negligible at high temperatures. The same is true for the mirror sector except it may have a lower temperature $T' < T$ (e.g., $T'=1/3T$) at the same time as suggested by previous studies \cite{kolb1985,hodges1993,foot2004,berezhiani2004}. In fact, lower mirror temperature can occur naturally after inflation and subsequent reheating \cite{kolb1985,hodges1993,berezhiani2006} and it requires $T' < T/2$ to be consistent with BBN, in particular, the observed primordial helium abundance \cite{kolb1985,hodges1993}. Such a standard mirror temperature condition ($T'/T < 1/2$) is sufficient for the following discussions.

Oscillations of $n-n'$ then become the dominant source for matter exchange between the two parallel sectors as other neutral particles are either too short-lived (e.g., $\pi^0$) or too light (like neutrinos) to contribute, which will be discussed later. Therefore, the baryon contents of the two sectors will evolve via the interplay of $n-n'$ oscillations as follows,
\begin{eqnarray} \label{eq_move}
\frac{d\chi(t)}{dt} &=& \frac{1}{2}P_{n'n}(\tau_f') \lambda_{np}'(t) \chi'(t) - \frac{1}{2}P_{nn'}(\tau_f) \lambda_{np}(t) \chi(t), \\
\label{eq_move2}
\frac{d\chi'(t)}{dt} &=& \frac{1}{2}P_{nn'}(\tau_f) \lambda_{np}(t) \chi(t) - \frac{1}{2}P_{n'n}(\tau_f') \lambda_{np}'(t) \chi'(t)
\end{eqnarray}
where $P_{nn'}$ ($P_{n'n}$) is the same as defined in Eq. (\ref{eq_prob}) with $t=\tau_f(\tau_f')$. The conversion rate $\lambda_{np}$ between protons and neutrons (smaller $n-\pi$ contributions are ignored here \cite{bringmann2019}) essentially defines the mean free flight time $\tau_f$ as \cite{weinberg1972},
\begin{eqnarray}
\frac{1}{\tau_f} = \lambda_{np} &=& \frac{7\pi}{30}G_F^2 |V_{ud}|^2 (1+3\left( \frac{g_A}{g_V}\right)^2)(kT)^5 \nonumber \\
 &\sim & 0.4 T_{10}^5 \text{[sec$^{-1}$]}
\end{eqnarray}
where $G_F$ is the Fermi constant, $V_{ud}$ is the CKM matrix element, and $g_A/g_V$ is the ratio of axial-vector/vector couplings.

Under the condition of the lower mirror temperature discussed above, the two equations (\ref{eq_move}-\ref{eq_move2}) will be decoupled and can be simplified by removing the first term. Therefore, the matter exchange will be in two separate steps. First, mirror neutrons, formed earlier than ordinary neutrons, will be converted to neutrons and hence mirror matter to ordinary matter due to $n-p$ equilibrium. The second step starts when the ordinary temperature gets low enough so that ordinary neutrons/matter will go back to the pool of mirror matter in the same way. The second step is much more significant due to a slower universe expansion rate at a later time as detailed below. In the end, a small amount of ordinary matter (neutrons and protons) is left while the mirror matter dominates the universe behaving exactly like the dark matter we have observed today.

\begin{figure}
\includegraphics[scale=0.65]{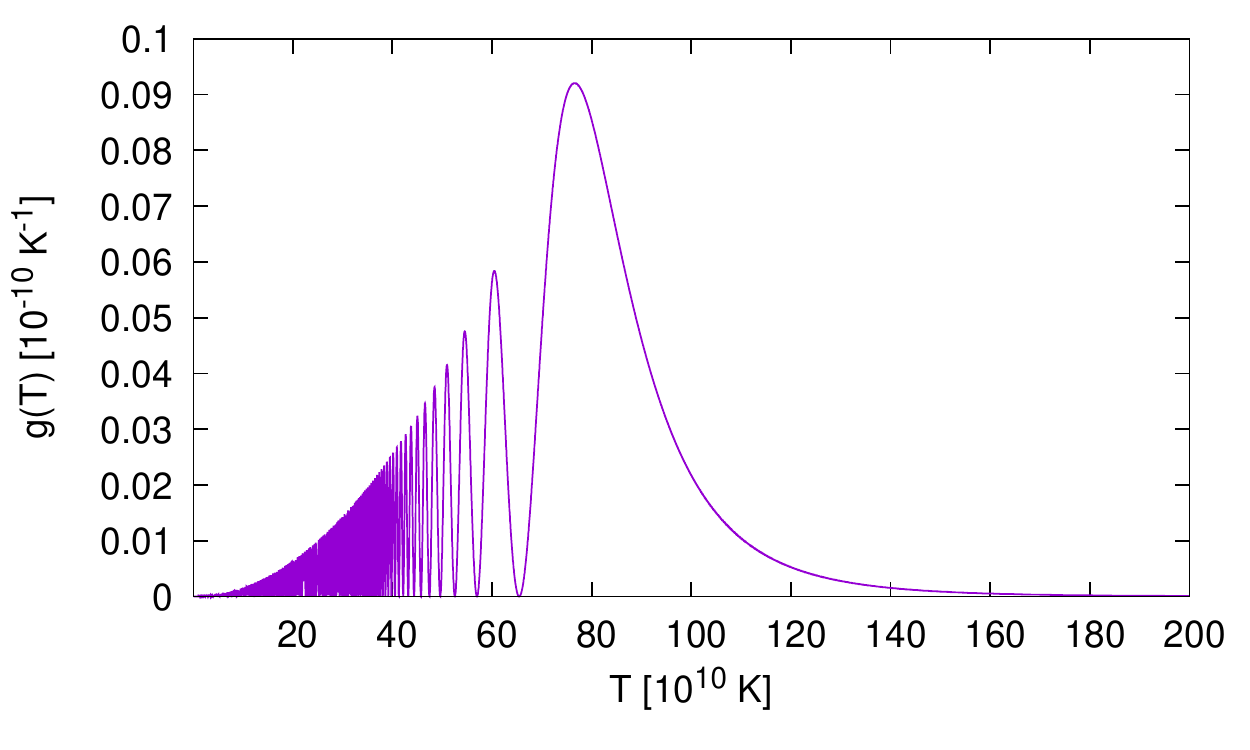}
\caption{\label{fig_1} The temperature dependence of the $n \leftrightarrow n'$ conversion factor $g(T)$ due to $n-n'$ oscillation is shown. The peak conversion rate occurs right below $T=10^{12}$ K shortly after (mirror) baryons are formed in the early universe.}
\end{figure}

First, one can examine the last yet dominant $n \rightarrow n'$ conversion process. The fraction of leftover ordinary matter can be worked out as follows,
\begin{equation} \label{eq_ratio}
\frac{\chi_r}{\chi_i} = \exp(-\frac{1}{2}\int P_{nn'}(\tau_f) \lambda_{np}(t)dt)
\end{equation}
where $\chi_r$ ($\chi_i$) is the remaining (initial) amount of ordinary matter. The integration over time in Eq. (\ref{eq_ratio}) can be simplified by replacing $t$ with temperature using Eq. (\ref{eq_age}),
\begin{eqnarray} 
\int g(T)dT &\equiv& \int P_{nn'} \lambda_{np}dt \nonumber \\
\label{eq_int}
&=& \int 1.6\times10^{-5} \sin^2(\frac{\Delta_{nn'}/\text{[eV]} }{5.3\times10^{-16} (T_{10})^5}) (T_{10})^2 dT_{10}
\end{eqnarray}
where the conversion factor $g(T)$ is plotted in Fig. \ref{fig_1} assuming the $n-n'$ mass difference $\Delta_{nn'} = 2\times10^{-6}$ eV and the mixing strength of $2\times 10^{-5}$. As seen in Fig. \ref{fig_1}, the peak conversion occurs just under $10^{12}$ K (i.e., at $\sim 70$ MeV) and the distribution is narrow enough to decouple the evolution equations. Similar equations as above also apply to the first or $n' \rightarrow n$ conversion step although it is much shorter and the conversion factor is greatly suppressed by the small factor of $(T'/T)^2\sqrt{g_*(T')/g_*(T)}$. Therefore, the contribution from the $n' \rightarrow n$ conversion step is negligible.

For $\Delta_{nn'} \sim 2\times10^{-6}$ eV, after the conversion ($n \rightarrow n'$) process following Eqs. (\ref{eq_ratio}-\ref{eq_int}) over the temperature range between the QCD phase transition ($T_c=150-200$ MeV) and the weak interaction decoupling ($T=1$ MeV), most of the ordinary matter is converted to mirror matter resulting a mirror-to-ordinary matter ratio of about 5.4, which is the same as the ratio of dark matter to baryon matter. As it turns out, the obtained $\Delta_{nn'}$ value within $50\%$ is very insensitive to other parameters such as the phase transition or nucleon-forming temperature (e.g., between 150 and 200 MeV) and the initial mirror-to-ordinary baryon ratio (e.g., equal amounts or little initial net mirror baryon matter as suggested by a separate work \cite{tan2019c}). Remarkably, $\Delta_{nn'} \sim 2\times10^{-6}$ eV is consistent with constraints from the neutron lifetime experiments as discussed above. A high-precision laboratory measurement of this mass splitting parameter is proposed using a ``beam'' approach with very high magnetic fields \cite{tan2019d}.

Conversely, using the observed dark-to-baryon matter ratio as a constraint, one can obtain the following simple relationship between the $n-n'$ mass difference and its mixing strength,
\begin{equation}
\sin^2(2\theta) = \left( \frac{3\times 10^{-14} \text{eV}}{\Delta_{nn'}}\right)^{0.6}
\end{equation}
which could be used to determine a better mass difference once the UCN experiments have better measurements for the mixing strength.

If the $n-n'$ mixing strength is on the order of $10^{-5}-10^{-6}$, the corresponding single quark mixing strength will be the cube root of that, i.e., about $10^{-2}$. The neutral mesons like $\pi^0$ and $K^0$, consequently, will have a mirror mixing strength of about $10^{-4}$. The mixing probabilities for neutral mesons are,

\begin{eqnarray}
P_{\pi^0\pi^{0'}}(t) & = & \sin^2(2\theta_{\pi^0}) \sin^2(\frac{1}{2}\Delta_{\pi^0\pi^{0'}} t), \nonumber \\
\label{eq_probmeson}
P_{K^0K^{0'}}(t) &=& \sin^2(2\theta_{K^0}) \sin^2(\frac{1}{2}\Delta_{K^0K^{0'}} t)
\end{eqnarray}
which hold true even for relativistic particles as far as $t$ is the proper time in the particle's rest frame. The mirror particles are not detectable in the ordinary world so that Eq. (\ref{eq_probmeson}) essentially defines the branching fractions of invisible decays of the mesons. Since the mass difference stems from the Higgs mixing, it is reasonable to assume that it is scaled to the particle's mass. Therefore, $\Delta_{\pi^0\pi^{0'}}$ and $\Delta_{K^0K^{0'}}$ should be similar to that of $n-n'$, i.e., about $10^{-6}$ eV. Considering the $\pi^0$'s very short lifetime of $8.52\times10^{-17}$ s, the $\pi^{0} - \pi^{0'}$ transition probability or the branching fraction of its invisible decays should be less than $10^{-18}$ which is not detectable with today's technology. On the other hand, $K^0$ has fairly long lifetime ($9\times10^{-11}$ s for $K^0_S$ and $5\times10^{-8}$ s for $K^0_L$) which makes the propagation factor in Eq. (\ref{eq_probmeson}) about $10^{-2}$ for $K^0_S$ and averaged to 1/2 for $K^0_L$. Therefore, the branching fraction of $K^0$ invisible decays is estimated to be about $10^{-6}$ for $K^0_S$ and $10^{-4}$ for $K^0_L$, which surprisingly is not constrained experimentally \cite{gninenko2015}. Such a large fraction should motivate people to start searching for $K^0 \rightarrow invisible$ decays at current kaon production facilities.

Similar estimate can be done for $D^0$ and $B^0$ mesons and their lifetimes permit an invisible branching fraction of about $10^{-9}-10^{-10}$ from the mirror oscillations. Other heavy neutral particles have even shorter lifetimes so that the effect of the oscillations is negligible. As for the light particles, photons have no rest mass and thus can not be mixed. The massive species of neutrinos should take part in the mirror mixing just like the 3-generation mixing in the ordinary sector. However, the effect is very small as $\Delta^2_{\nu\nu'} \sim 10^{-17} - 10^{-19}$ eV$^2$ assuming a neutrino mass of $0.1-0.01$ eV. To observe this oscillation effect for 1 MeV neutrinos, it has to come from stars at least thousands of light years away, possibly from a supernova explosion. Solar neutrinos have to have an energy below 1 eV to experience such oscillations on its way to Earth. 

To conclude, the following analysis and experimental studies are highly recommended in order to test the proposed model.
Careful reanalysis of past ``bottle'' experiments should be carried out by taking into account the mean free flight time $\tau_f$ evaluated or simulated for its own specific setup. Under this model with Eq. (\ref{eq_prob3}) for the corrections from $n-n'$ oscillations, a consistent beta decay lifetime should be obtained and it will also help determine a more accurate $n-n'$ mixing strength. Magnetic traps with various sizes can provide a much stricter test of this model without worries of the interference from wall surface. Studies of $K^0\rightarrow invisible$ decays should be granted high priority at kaon production facilities. The measured invisible branching fraction will tell us about the $K^0-K^{0'}$ mixing strength and possibly verify the mechanism of the spontaneously broken mirror symmetry. If this mirror symmetry theory is confirmed, invisible mirror stars and galaxies should be searched. Such candidates may have already been observed in most of the black hole and neutron star merger events that were detected by gravitational-wave observatories but could not be identified with its electromagnetic counterpart except for the one neutron star merger \cite{ligoscientificcollaborationandvirgocollaboration2017}. Could most of the merger events actually come from the mirror sector of the universe? This is understandable since we are in a dark (mirror) matter dominated universe.

Many of the intriguing features conceived in previous studies of the mirror matter theory \cite{blinnikov1983,kolb1985,khlopov1991,foot2004,berezhiani2004} are kept and work even better under the new model. For example, the  $\Omega_{dark}/\Omega_B$ ratio could be explained better as discussed above. Another example is the unexpected excess of ultrahigh-energy cosmic rays above the Greisen–Zatsepin–Kuzmin (GZK) limit and an explanation using the mirror matter theory was provided except for a caveat of unrealistic requirement on galactic and intergalactic magnetic fields \cite{berezhiani2006a}. Under the current model, better explanation without tarnishment for this and various other GZK related effects is presented in a separate work \cite{tan2019b}. To resolve the galactic dark matter issues, a requirement of strongly self-interacting dark matter \cite{spergel2000} was proposed and it can be naturally met with the mirror matter theory.

Based on this model, the Standard Model is extended with mirror matter and used for understanding dark energy and puzzles in particle physics \cite{tan2019e}. Application of this proposed $n-n'$ model to evolution and nucleosynthesis in stars is studied under a new stellar burning theory \cite{tan2019a}. Remarkable agreement between the observations and the predictions from the study provides strong evidence and support for this model \cite{tan2019a}. And furthermore, a natural extension of the new model applying kaon oscillations in the early universe shows a promising solution to the long-standing baryon asymmetry problem with new insights for the QCD phase transition and B-violation topological processes \cite{tan2019c}. Last but not least, extension of the CKM matrix and laboratory tests of the new model are proposed in a separate work \cite{tan2019d}.

The influence of this $n-n'$ mixing model can also be studied in various other scenarios like BBN where the $^7$Li problem could potentially be solved \cite{coc2013,coc2014a,tan2019c}, stellar burning processes (in particular, neutron capture processes) \cite{tan2019a}, neutron star mergers (including all three cases of ordinary-ordinary, mirror-mirror, and mirror-ordinary mergers). Probably the two mirrored yet separated worlds have been and are being connected by the active and fascinating messenger of the $n-n'$ doublet during the Big Bang and after the formation of stars.

\section*{Acknowledgements}
I would like to thank Ani Aprahamian and Michael Wiescher for supporting me in a great research environment at Notre Dame. Useful discussions on UCN with Adam Holley at the 42nd Symposium on Nuclear Physics in Mexico are appreciated. I also like to thank Jim Cline for informing about their work on the effect of pion-neutron interactions under a different neutron oscillation mechanism, Bo Feng and Jing Shu for pointing out the connection of this work to the twin Higgs models.
This work is supported in part by the National Science Foundation under
grant No. PHY-1713857 and the Joint Institute for Nuclear Astrophysics (JINA-CEE, www.jinaweb.org), NSF-PFC under grant No. PHY-1430152.



\bibliographystyle{elsarticle-num}
\bibliography{nosc}





\end{document}